\documentclass[10pt,twocolumn,letterpaper]{article}
\usepackage[pagenumbers]{cvpr} 
\usepackage[accsupp]{axessibility} 

\usepackage[dvipsnames]{xcolor}

\usepackage{colortbl}

\definecolor{tablegray}{gray}{.9}
\definecolor{tableblue}{RGB}{221,238,251}
\definecolor{tablegreen}{RGB}{225,245,245}
\definecolor{baselinecolor}{gray}{.9}
\newcommand{\baseline}[1]{\cellcolor{baselinecolor}{#1}}

\newcolumntype{x}[1]{>{\centering\arraybackslash}p{#1pt}}
\newcolumntype{y}[1]{>{\raggedright\arraybackslash}p{#1pt}}
\newcolumntype{z}[1]{>{\raggedleft\arraybackslash}p{#1pt}}

\def\eg{\emph{e.g}\onedot} 
\def\ie{\emph{i.e}\onedot}

\def\etal{\emph{et al}\onedot}

\newlength\savedwidth
\newcommand\whline{\noalign{\global\savedwidth\arrayrulewidth\global\arrayrulewidth 1.0pt}
\hline\noalign{\global\arrayrulewidth\savedwidth}}

\newcolumntype{"}{@{\hskip\tabcolsep\vrule width 1pt\hskip\tabcolsep}}

\newcommand{\tablestyle}[2]{\setlength{\tabcolsep}{#1}\renewcommand{\arraystretch}{#2}\centering\footnotesize}

\usepackage{amsfonts}
\usepackage{algorithmic}
\usepackage{algorithm}
\usepackage{array}
\usepackage{textcomp}
\usepackage{stfloats}
\usepackage{url}
\usepackage{verbatim}
\usepackage{color}
\usepackage{multirow}
\usepackage{bbm}
\usepackage{bm}
\usepackage{makecell}

\definecolor{cvprblue}{rgb}{0.21,0.49,0.74}
\usepackage[pagebackref,breaklinks,colorlinks,citecolor=cvprblue]{hyperref}
\def\paperID{9818} 
\def\confName{CVPR}
\def\confYear{2024}

\title{Modular Blind Video Quality Assessment}

\author{Wen Wen$^{1}$, Mu Li$^{2}$, Yabin Zhang$^{3}$, Yiting Liao$^{3}$, Junlin Li$^{3}$, Li Zhang$^{3}$, and Kede Ma$^{1}$\thanks{Corresponding author.}\\
$^1$ City University of Hong Kong, 
$^2$ The Chinese University of Hong Kong, Shenzhen\\
$^3$ ByteDance Inc.\\
{\tt  wwen29-c@my.cityu.edu.hk, limuhit@gmail.com}\\
{\tt \{zhangtao.ceb, liaoyiting, lijunlin.li, lizhang.idm\}@bytedance.com}\\
{\tt kede.ma@cityu.edu.hk}\\
\url{https://github.com/winwinwenwen77/ModularBVQA}
}

\begin{document}
\maketitle

\begin{abstract}
Blind video quality assessment (BVQA) plays a pivotal role in evaluating and improving the viewing experience of end-users across a wide range of video-based platforms and services. Contemporary deep learning-based models primarily analyze video content in its aggressively subsampled format, while being blind to the impact of the actual spatial resolution and frame rate on video quality.
In this paper, we propose a modular BVQA model and a method of training it to improve its modularity. Our model comprises a base quality predictor, a spatial rectifier, and a temporal rectifier, responding to the visual content and distortion, spatial resolution, and frame rate changes on video quality, respectively. During training, spatial and temporal rectifiers are dropped out with some probabilities to render
the base quality predictor a standalone BVQA model, which should work better with the rectifiers. Extensive experiments on both professionally-generated content and user-generated content video databases show that our quality model achieves superior or comparable performance to current methods. Additionally, the modularity of our model
offers an opportunity to analyze existing video quality databases in terms of their spatial and temporal complexity. 

\end{abstract}

\section{Introduction}
\label{sec:intro}

We undoubtedly reside in an era that exposes us to a diverse array of exponentially growing video data on a daily basis. Such growth is accompanied by advancements in video recording equipment and display devices, leading to high spatial resolution, high frame rate, high dynamic range, and wide color gamut video content. As a result, understanding how multifaceted video attributes together affect video quality is crucial to measure and improve the viewing experience of end-users. Over the years, researchers have collected numerous supporting evidence from psychophysical and perceptual studies~\cite{zhai2008cross,janowski2010qoe,nasiri2015perceptual,li2019avc,ying2021patch,mackin2018studyfr,madhusudana2021subjective} that a higher spatial resolution and higher frame rate contribute positively to video quality, with exact perceptual gains depending on the video content.

In response to these subjective findings, early knowledge-driven BVQA models directly take spatial resolution and frame rate parameters as part of input for quality prediction of compressed videos~\cite{ou2014q}. 
Despite the simplicity, these video attribute parameters, being independent of content and distortion, are less relevant to perceived video quality.
Consequently, simple content complexity features like spatial information\footnote{Spatial information is computed as the root mean squares of the gradient magnitudes of each frame.} and temporal information\footnote{Temporal information is computed as the root mean squares of the differences of consecutive frames.}, along with simple natural scene statistics derived from the pixel domain~\cite{mittal2015completely}, the 3D discrete cosine transform domain~\cite{saad2014blind}, and the wavelet domain~\cite{dendi2020no} are computed at the actual spatial resolution and frame rate~\cite{janowski2010qoe}.
Nevertheless, operating on the full-size video sequence is computationally daunting. Korhonen~\cite{korhonen2019two} proposed a two-level approach, extracting low-complexity features from each spatially downsampled frame and high-complexity features from key frames at the actual spatial resolution. 
An alternative approach is to work with spatiotemporal chips~\cite{ebenezer2021chipqa}, which are spatial-time localized cuts of the full-size video in local motion flow directions. In general, knowledge-driven BVQA models perform marginally due to the limited expressiveness of manually crafted features.

\begin{figure*}[!t]
\scriptsize
\centering
\includegraphics[width=1.0\textwidth]{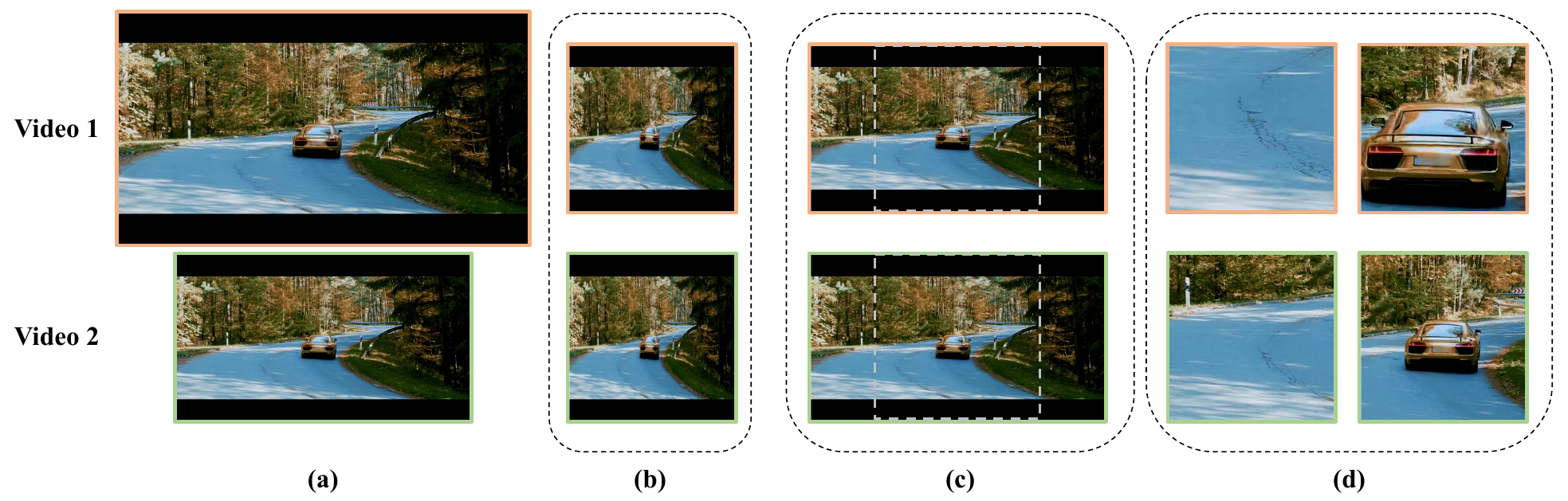}
\caption{Conventional ways of pre-processing videos in spatial view. \textbf{(a)} Videos with the same content but different spatial resolutions taken from the Waterloo-IVC-4K dataset~\cite{li2019avc}. \textbf{(b)} Resizing without maintaining the aspect ratio leads to geometric distortions of structures and textures. \textbf{(c)} Aspect ratio-preserving resizing and cropping results in almost identical inputs of fixed size. 
\textbf{(d)} Cropping from videos at the actual spatial resolution reduces the field of view with limited content coverage.}
\vspace{-12pt}
\label{fig:spatial}
\end{figure*}

\begin{figure}[!tbp]
\scriptsize
\centering
\includegraphics[width=0.47\textwidth]{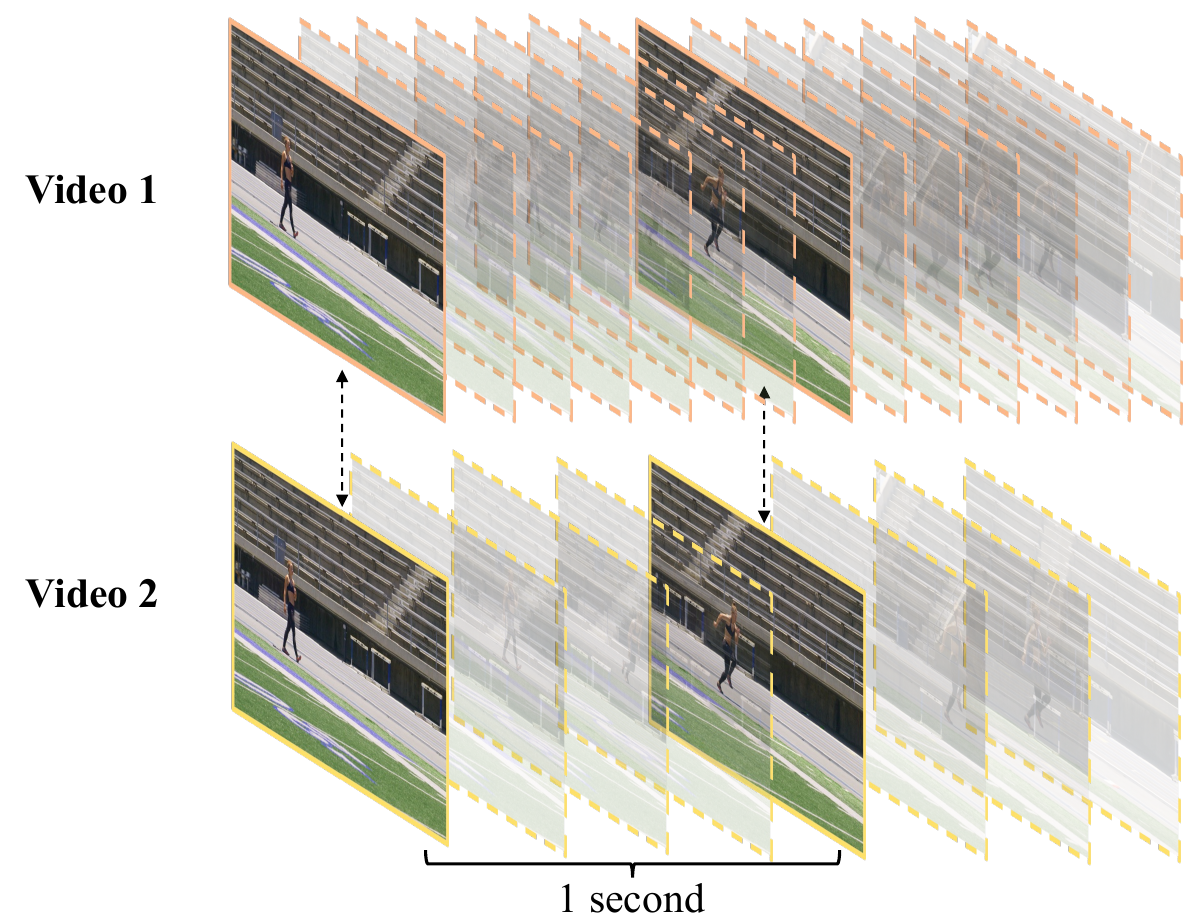}
\caption{Two videos from the LIVE-YT-HFR dataset~\cite{madhusudana2021subjective} with identical content and duration but different frame rates. When the subsampling rate is proportional to the frame rate, the remaining frames are identical.}
\vspace{-12pt}
\label{fig:temporal}
\end{figure}

The computational issues faced by deep learning-based data-driven BVQA methods are more pronounced. There are few attempts~\cite{li2019quality,li2022blindly} to assess full-size videos, which come with significant computational demands, especially for videos of high resolutions and frame rates. Moreover, due to the small scale of video quality datasets, many BVQA methods that utilize convolutional neural networks (CNNs) commonly depend on pretrained models from object recognition tasks, which expect small and fixed-size inputs. Consequently, videos need to be spatially resized~\cite{yi2021attention, wang2021rich, sun2022deep} and/or cropped~\cite{ying2021patch, wang2021rich, wu2022fast, wu2022disentangling}, and temporally subsampled~\cite{yi2021attention, sun2022deep, wang2021rich}.
When two videos of the same scene but different spatial resolutions are resized to the same lower resolution, their quality variations, which are discernable by the human eye, diminish (see Figure~\ref{fig:spatial} (b) and (c)).  Cropping, on the other hand, significantly reduces the field of view and content coverage, which can impede the precision of BVQA. When videos that share the same content and length but differ in frame rates undergo temporal subsampling at a rate proportional to their frame rates,  the frames that remain would be identical, as shown in Figure~\ref{fig:temporal}.
Therefore, these BVQA methods are insensitive to changes in the spatial resolution and frame rate and their impact on video quality.

To reliably assess the perceptual quality of digital videos with great content and distortion diversities, and variable spatial resolutions and frame rates, we propose a modular BVQA model. Our model consists of three modules: a base quality predictor, a spatial rectifier, and a temporal rectifier, responding to the visual content and distortion, spatial resolution, and frame rate changes, respectively. The base quality predictor takes a sparse set of spatially downsampled key frames as input, and produces a scalar as the quality estimate. The spatial rectifier relies on a shallow CNN to process the Laplacian pyramids of the key frames at the actual spatial resolution, and computes the scaling and shift parameters to rectify the base quality score. Similarly, the temporal rectifier relies on a lightweight CNN to process the spatially downsampled video chunks centered around the key frames at the actual frame rate, and computes another scaling and shift parameters for quality rectification. To enhance the modularity of our model, we introduce a dropout strategy during training. At each iteration, we randomly drop out the spatial and/or temporal rectifiers with pre-specified probabilities. 
In summary, the current work presents several novel elements:
\begin{itemize} 
\item A modular BVQA model that is sensitive to visual distortions and changes in the spatial resolution and frame rate, and is readily extensible to add other video attributes (\eg, dynamic range and color gamut) as additional rectifiers;
\item A training strategy for our BVQA model to encourage the base quality predictor to function independently as a BVQA model, which would perform better when combined with the rectifiers;
\item A experimental demonstration of the superiority of our BVQA model over existing methods on a total of \textit{fourteen} video quality datasets~\cite{mackin2018studyreso,li2019avc,mackin2018studyfr,madhusudana2021subjective,shang2021assessment, lee2021subjective, nuutinen2016cvd2014, ghadiyaram2017capture,sinno2018large, hosu2017konstanz,wang2019youtube,chen2019qoe, yu2022subjective,ying2021patch}, covering professionally-generated content (PGC) and user-generated content (UGC), synthetic and authentic distortions, and  variable spatial resolutions and frame rates;
\item A computational analysis of eight UGC datasets in terms of their spatial and temporal complexity thanks to the modularity design of our BVQA model~\cite{sun2024analysis}.
\end{itemize}

\section{Related Work}
In this section, we provide a concise overview of BVQA models for PGC and UGC content.
\label{sec:related}
\subsection{BVQA Models for PGC Content}

Several publicly available PGC datasets~\cite{mackin2018studyreso,li2019avc,mackin2018studyfr,madhusudana2021subjective,rao2019avt,lee2021subjective} have been specifically constructed to investigate the effects of spatial resolution and frame rate variations on video quality. However, there is a lack of research dedicated to BVQA methods for various spatial resolution and frame rate configurations.
Ou~\etal~\cite{ou2010modeling} designed an exponential function to compute video quality, taking into account the combined effect of a spatial quality factor (\eg, the peak signal-to-noise ratio (PSNR)) and a temporal quality factor (\eg, the frame rate).
Janowski and Romaniak~\cite{janowski2010qoe} proposed a generalized linear function to model spatial and temporal information independently. 
Ou~\etal~\cite{ou2014q} developed Q-STAR, a model that measures the joint influence of the spatial resolution, temporal resolution, and quantization step size through the product of three one-parameter functions. 
FAVER~\cite{zheng2022no} was designed to evaluate videos with diverse and high frame rates. It extracts natural scene statistics from the spatiotemporal wavelet domain, and employs a support vector regressor to estimate video quality at varying frame rates.

\subsection{BVQA Models for UGC Content}
UGC videos cover a wide range of natural scenes,  spatial resolutions, and frame rates, leading to a diverse array of spatiotemporal distortions. Current BVQA models have achieved considerable progress in assessing the quality of UGC videos. In particular, knowledge-driven BVQA models primarily rely on manually crafted features from both spatial and temporal domains to determine video quality. Notable examples of such models are V-BLIINDS~\cite{saad2014blind}, VIIDEO~\cite{mittal2015completely}, TLVQM~\cite{korhonen2019two}, and VIDEVAL~\cite{tu2021ugc}.
Recent studies have been focusing on enhancing BVQA models by integrating handcrafted features with those computed by pretrained CNNs. Representative models in this regard include CNN-TLVQM~\cite{korhonen2020blind} and RAPIQUE~\cite{tu2021rapique}. Meanwhile, there has been a shift towards purely learning-based approaches.  VSFA~\cite{li2019quality}, Li22~\cite{li2022blindly}, PVQ~\cite{ying2021patch}, and CoINVQ~\cite{wang2021rich} leverage various pretrained networks as fixed feature extractors, and train the regression module independently. Yi21~\cite{yi2021attention} and SimpleVQA~\cite{sun2022deep} reduce the video resolution in space and time,
followed by end-to-end fine-tuning. FastVQA~\cite{wu2022fast} extracts spatially localized and temporally aligned cubes (termed as fragments) to facilitate end-to-end training. DOVER~\cite{wu2022disentangling} further enhances FastVQA by integrating a pretrained ConvNeXt-T~\cite{liu2022convnet} designed for image aesthetics assessment.

\section{Proposed Method}
\label{sec:prop}
\begin{figure*}
\scriptsize
\centering
\includegraphics[width=1\textwidth]{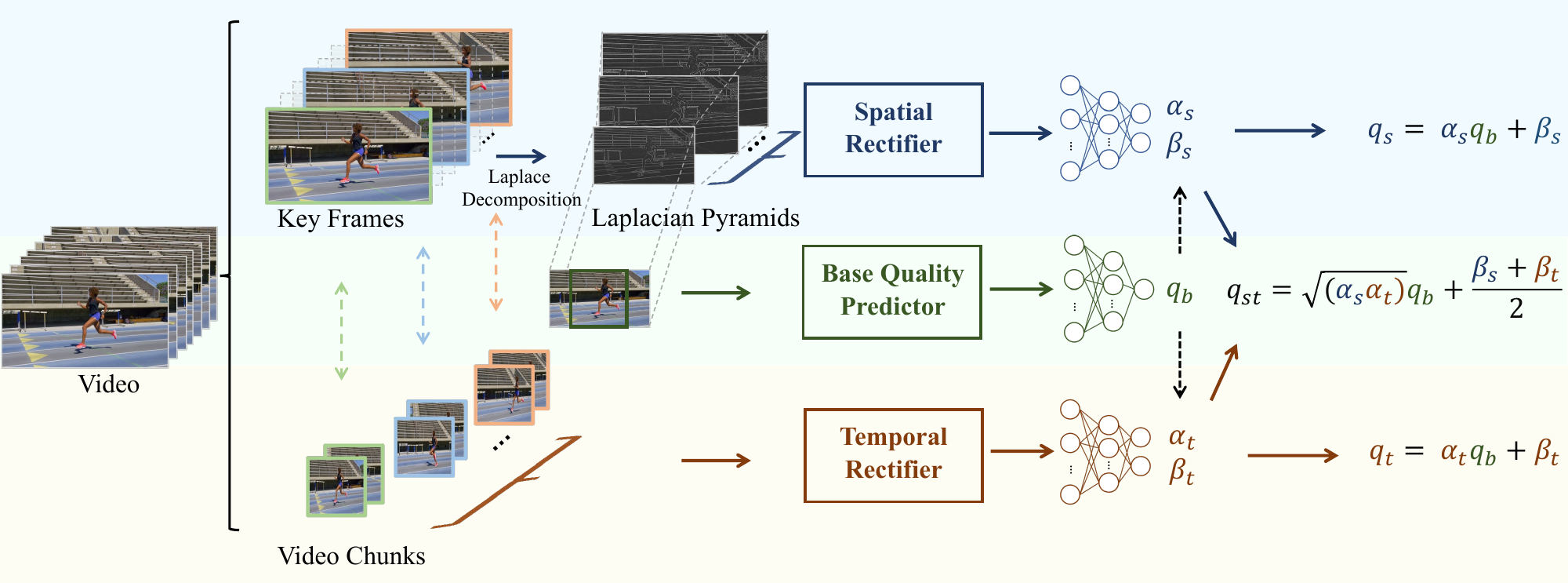}
\caption{System diagram of our modular BVQA model. The base quality predictor takes a sparse set of spatially downsampled key frames as input, and generates a base quality score denoted by $q_b$. The spatial rectifier employs Laplacian pyramids derived from the key frames at their actual spatial resolution, and computes a scaling parameter $\alpha_s$ and a shift parameter $\beta_s$ to rectify the base quality score.
The temporal rectifier leverages features from the video chunks centered around the key frames at the actual frame rate to compute another scaling parameter $\alpha_t$ and shift parameter $\beta_t$ for quality rectification. 
}
\label{fig:model_structure}
\vspace{-12pt}
\end{figure*}

In this section, we present in detail our modular BVQA model, comprising a base quality predictor, a spatial rectifier, and a temporal rectifier, as illustrated in Figure~\ref{fig:model_structure}.

\subsection{Base Quality Predictor}
We denote a video sequence as $\bm{x}= \{\bm x_i\}_{i=0}^{N-1}$, where $\bm x_i\in\mathbb{R}^{H\times W \times 3}$ represents the $i$-th frame,  $H$ and $W$ are the frame height and width, and $N$ is the total number of frames. The base quality predictor $f(\cdot):\mathbb{R}^{H\times W\times 3\times N}\mapsto \mathbb{R}$ takes $\bm {x}$ as input, and computes a quality  score, $q_b$. As part of the modular design, $f(\cdot)$ responds to semantic visual content and spatial distortions that remain unaffected by spatial resizing. Specifically, we uniformly sample a sparse set of $M$ key frames, $\bm y = \{\bm y_i\}_{i=0}^{M-1}$, which are subject to bicubic resizing and cropping to $H_b\times W_b$. Other parametric temporal subsampling techniques~\cite{korbar2019scsampler} are also applicable with added computational complexity. $H_b$ and $W_b$ are determined by the input specifications of the pretrained network, which in our paper is a vision Transformer (ViT)~\cite{dosovitskiy2020image}  in the CLIP model~\cite{radford2021learning}. We use the image representation corresponding to the \texttt{[class]} token, and add a multilayer perceptron (MLP) with two layers (\ie, two fully connected layers with ReLU nonlinearity in between) as the quality regressor. We compute per-key-frame quality scores, and average them to obtain a base quality estimate of $\bm x$.

\subsection{Spatial Rectifier}

We incorporate a spatial rectifier, $\bm{f}_s(\cdot):\mathbb{R}^{H \times W \times 3\times N} \mapsto \mathbb{R}^{2 \times 1}$, that takes the same $\bm x$ as input and computes a scale parameter $\alpha_s$ and a shift parameter $\beta_s$ to rectify the base quality score:
\begin{align}\label{eq:sr}
q_{s} = \alpha_s q_b+\beta_s.
\end{align}
As part of the modular design, $\bm f_s(\cdot)$ responds to spatial distortions that arise from or are affected by spatial resizing. Similar to the base quality predictor, we work with the sparse set of $M$ key frames, $\bm y = \{\bm y_i\}_{i=0}^{M-1}$. The difference lies in that we do not further perform spatial resizing, but build a Laplacian pyramid of $K+1$ levels for each key frame at the actual spatial resolution. For the $i$-th key frame and at the $k$-th level, we have
\begin{align}
    \bm y_i^{(k+1)} &= \mathbf{D}\mathbf{L}\bm y_i^{(k)},\quad k\in\{0, \ldots, K-1\},\\
    \bm z_i^{(k)} &= \bm y_i^{(k)} - \mathbf{L}\mathbf{U}\bm y_i^{(k+1)},\\
    \bm z_i^{(K)} &= \bm y_i^{(K)},
\end{align}
where $\bm y_i^{(0)}  = \bm y_i$ and $\bm z^{(k)}_i$ is the $k$-th bandpass subband. $\mathbf{D}$ and $\mathbf{U}$ denote linear sub/upsampling operation, respectively, and $\mathbf{L}$ denotes the lowpass filter in the matrix form. We set the subsampling factor $\rho$ at each level to $\frac{ \min\{H, W\}}{\min\{H_b, W_b\}\times K}$. In case $\rho$ is a non-integer, we implement sub/upsampling together with the lowpass filtering using bicubic interpolation. The $K$-th level of the Laplacian pyramids $\bm z^{(K)}_i$ represents a low-frequency residual, which is related to the input of the base quality predictor by a cropping operator, and can thus be discarded. 

We upsample all bandpass subbands to the actual resolution using bicubic interpolation, and feed each subband independently to a pretrained lightweight CNN for low-level feature extraction, followed by concatenation of $M\times K$ sets of spatial features. In our paper, we utilize the first two stages of ResNet-18~\cite{He2016ResNet} pretrained on ImageNet-1K~\cite{deng2009imagenet}. We last aggregate spatial information using global average and standard deviation pooling, and append a two-layer MLP to generate the spatial scaling parameter $\alpha_s$ and shift parameter $\beta_s$ in Eq.~\eqref{eq:sr}.

\subsection{Temporal Rectifier} 
We include a temporal rectifier $
\bm f_t(\cdot):\mathbb{R}^{H\times W\times 3\times N\mapsto \mathbb{R}^{2\times1}}$, that takes the same $\bm x $ as input and computes another scaling parameter $\alpha_t$ and  shift parameter $\beta_t$ to rectify the base quality score:
\begin{align}\label{eq:tr}
q_t = \alpha_t q_b + \beta_t.
\end{align}
As part of the modular design, $\bm{f}_t(\cdot)$ responds to temporal distortions resulting from motion anomaly and frame rate variations.
We sample a set of $M$ video chunks, denoted by $\bm v = \{\bm v_i\}_{i=0}^{M-1}$ from $\bm{x}$. Each $\bm{v}_i$ is centered around the $i$-th key frame $\bm{y}_i$, and consists of $L$ consecutive frames,
 resized to $H_v \times W_v$ without respecting the aspect ratio.
We utilize the fast pathway of a pretrained SlowFast network~\cite{feichtenhofer2019slowfast}, excluding the original classification head. Temporal features for different video chunks are concatenated, spatially pooled, and fed to a two-layer MLP to produce the temporal scaling parameter $\alpha_t$ and shift parameter $\beta_t$ in Eq.~\eqref{eq:tr}.

\subsection{Module Aggregation} 
During training, we implement a dropout strategy by randomly disabling the spatial and temporal rectifiers with some probabilities $p_s$ and $p_t$, respectively.
We derive the combined scaling and  shift parameters as the geometric and arithmetic mean of those from the activated rectifiers:
\begin{align}\label{eq:alpha_st}
\alpha_{st} = \left(\alpha_s^{\mathbbm{I}[u_s \geq p_s]} \alpha_t^{\mathbbm{I}[u_t \geq p_t]}\right)^{\frac{1}{ \mathrm{max}\{1, \mathbbm{I}[u_s \geq p_s] + \mathbbm{I}[u_t \geq p_t]\} }},
\end{align}
and 
\begin{align}
    \beta_{st} = \frac{\mathbbm{I}[u_s \geq p_s]\beta_s + \mathbbm{I}[u_t \geq p_t]\beta_t}{ \mathrm{max}\{1, \mathbbm{I}[u_s \geq p_s] + \mathbbm{I}[u_t \geq p_t]\}},
\end{align}
where $\mathbbm{I}[\cdot]$ represents the indicator function, $u_s$ and $u_t$ are two samples drawn from the uniform distribution $\mathcal{U}[0,1]$. 
The final rectified quality score  becomes
\begin{align}\label{eq:sr_tr}
    q_{st} = \alpha_{st} q_b + \beta_{st}.
\end{align}
As specific instances, if both rectifiers are dropped, $q_{st} = q_b$, and  if both rectifiers are enabled:
\begin{align}\label{eq:2_sr_tr}
    q_{st} = \sqrt{(\alpha_s \alpha_t)} q_b + \frac{\beta_s +\beta_t}{2}.
\end{align}
Such a modular combination takes the contributions of the rectifiers into separate account, and is readily extensible to incorporate future rectifiers into the model design.

\section{Experiment}
In this section, we first describe the experimental setups. We then present the experimental results on the six PGC and eight UGC datasets, with emphasis on the computational analysis of the spatial and temporal complexity of UGC datasets. Last, we conduct a set of ablation experiments to complete our model analysis.

\begin{table*}\scriptsize
	\centering
	\renewcommand{\arraystretch}{1.25}
        \fontsize{8pt}{9pt}\selectfont
	\begin{tabular}{lcccccccc}
		Dataset & \# Videos & \# Scenes & Duration (Sec) & Spatial Resolution & Frame Rate & Distortion Type \\
	\whline
        BVI-SR~\cite{mackin2018studyreso} & 240 & 24 & 5 & 4K  & 60 & Subsampling \\
        Waterloo-IVC-4K~\cite{li2019avc} & 1,200 & 20 & 9-10 & 540p, 1080p, 4K  & 24, 25, 30 & Compression  \\
        BVI-HFR~\cite{mackin2018studyfr} & 88 & 22  & 10 & 1080p & 15, 30, 60, 120 & Frame Averaging  \\
        LIVE-YT-HFR~\cite{madhusudana2021subjective} & 480 & 16 & 6-10 & 1080p & 24, 30, 60, 82, 98, 120 & Compression  \\
        LIVE-Livestream~\cite{shang2021assessment} & 315 & 45 & 7 & 1080p, 4K & 25, 30 & Compression, Streaming \\
        ETRI~\cite{lee2021subjective} & 437 &  15 & 5-7  & 540p, 720p, 1080p, 4K & 30, 60, 120 &   Compression \\
        \hline
  	CVD2014~\cite{nuutinen2016cvd2014} & 234 & 5& 10-25& 480p,720p & 10-32 & Authentic   \\
        LIVE-Qualcomm~\cite{ghadiyaram2017capture} & 208 &54 & 15 & 1080p &  30& Authentic   \\
		KoNViD-1K~\cite{hosu2017konstanz} & 1,200 & 1,200 & 8 & 540p  & 24, 25, 30 & Authentic   \\
  	LIVE-VQC~\cite{sinno2018large} & 585 & 585 & 10 & 240p-1080p & 30 & Authentic \\
		YouTube-UGC~\cite{wang2019youtube} & 1,142 & 1,142  & 20 & 360p-4K & 30 & Authentic  \\
          LBVD~\cite{chen2019qoe} & 1,013 & 1,013 & 10 & 240p-540p  & $<$ 30 & Authentic, Streaming  \\
	LSVQ~\cite{ying2021patch} & 38,811 & 38,811 & 5-12 & 99p-4K  & $<$ 60 & Authentic \\
        LIVE-YT-Gaming~\cite{yu2022subjective} & 600& 600& 8-9 & 360p-1080p& 30, 60& Authentic, Rendering  \\
	\end{tabular}
\caption{Summary of PGC and UGC VQA datasets.}
\label{vqa_dataset}	
\vspace{-9pt}
\end{table*}

\subsection{Experimental Setups}
\noindent\textbf{Benchmark Datasets}. 
To assess the spatial rectifier, we employ BVI-SR~\cite{mackin2018studyreso} and Waterloo-IVC-4K~\cite{li2019avc}, which are dedicated to examining how spatial resolutions 
affect video quality.
To evaluate the temporal rectifier, we utilize BVI-HFR~\cite{mackin2018studyfr} and LIVE-YT-HFR~\cite{madhusudana2021subjective}, which are tailored to study the influence of frame rates on video quality. 
LIVE-Livestream and ETRI are selected to encompass video streaming-related artifacts.
We further validate the proposed BVQA model on eight UGC datasets~\cite{nuutinen2016cvd2014, ghadiyaram2017capture,sinno2018large, hosu2017konstanz,wang2019youtube,chen2019qoe, yu2022subjective,ying2021patch}. 
A comprehensive summary of these datasets is provided in Table~\ref{vqa_dataset}.

\noindent\textbf{Competing Methods}.
For PGC datasets, we choose six knowledge-driven BVQA models for comparison: 1) NIQE~\cite{Mittal2013NIQE}, 2) BRISQUE~\cite{mittal2012no}, 3) VIDEVAL~\cite{tu2021ugc}, 4) TLVQM~\cite{korhonen2019two}, 5) RAPIQUE~\cite{tu2021rapique} and 6) FAVER~\cite{zheng2022no}, which extract spatiotemporal features from full-size videos. FAVER~\cite{zheng2022no} is the only model capable of responding to different frame rates. We also include VSFA~\cite{li2019quality}, a learning-based model that takes full-size videos as input, and BTURA~\cite{lu2022deep}, which takes cropped frame patches as input.
For UGC datasets, we select 1) VSFA, 2) Li22~\cite{li2022blindly}, 3) SimpleVQA~\cite{sun2022deep}, 4) FastVQA~\cite{wu2022fast}, and 5) DOVER~\cite{wu2022disentangling} as competing models, among which FastVQA and DOVER deliver the state-of-the-art performance.

\noindent\textbf{Performance Criteria}.
We employ two evaluation criteria: the Spearman's rank correlation coefficient (SRCC) as a measure of prediction monotonicity and the Pearson linear correlation coefficient (PLCC) as a measure of prediction linearity. For the six PGC datasets, 
we adhere to the dataset splitting strategy described in the original papers, ensuring the content independence between the training and test sets. For the eight UGC datasets expect for LSVQ~\cite{ying2021patch}, we randomly split each into three non-overlapping subsets: $60\%$ for training, $20\%$ for validation, and $20\%$ for testing. All learning-based models are retrained in accordance with their original optimization procedures. For LSVQ, we conform to the training and testing protocol in the original paper. We repeat all random splitting processes $10$ times, and report the median results.

\noindent\textbf{Implementation Details}.
For the BVI-SR, Waterloo-IVC-4K, BVI-HFR, LIVE-YT-HFR, LIVE-Livestream, and ETRI datasets, we set the number of key frames $M$ to  $5$, $9$, $10$, $6$, $7$, and $5$, respectively. For the eight UGC datasets, $M = 8$. All key frames are first resized such that the shorter side has $224$ pixels and then cropped to $224 \times 224$.
For the spatial rectifier, the number of 
Laplacian pyramid levels $K$ is set to $5$, spanning from the actual spatial resolution down to $224$. 
Similarly, for the temporal rectifier, we extract an equal number of video chunks to that of the key frames.
During training, the spatial and temporal rectifiers are dropped out with a probability $p_s=p_t = 0.2$.
The optimization of all learnable parameters is carried out by the Adam method for $30$ epochs, with an initial learning rate of $1 \times 10^{-5}$, a decay ratio of $0.9$ per two epochs, and a minibatch size of $16$. We employ  PLCC as the optimization goal as it combines the advantages of both the regression and learning-to-rank formulations of BVQA.

\subsection{Results on PGC Datasets}

The proposed BVQA model is modular, and is thus capable of outputting four quality scores: $q_b$, $q_s$, $q_t$, and $q_{st}$. Table~\ref{performance_vaious_resolution} displays the results on BVI-SR and Waterloo-IVC-4K, which are predominantly influenced by the distortions arising from spatial resolution changes. Table~\ref{performance_vaious_framerate} shows the results on BVI-HFR and LIVE-YT-HFR, emphasizing the distortions arising from frame rate conversions. Table~\ref{performance_added_pgc} includes the results on LIVE-Livestream and ETRI, which encompass video streaming-based artifacts. By examining these results, we arrive at several interesting observations.

\begin{table}\footnotesize
\centering
\renewcommand{\arraystretch}{1.25}
\begin{tabular}{lcc}
 Method  & BVI-SR & Waterloo-IVC-4K\\
\whline
NIQE~\cite{Mittal2013NIQE}  &  0.494 / 0.590  & 0.099 / 0.160\\ 
BRISQUE~\cite{mittal2012no}  & 0.167 / 0.222 &  0.154 / 0.248 \\ 
VSFA~\cite{li2019quality}   & \textbf{0.831} / 0.780  & 0.122 / 0.148 \\
BTURA~\cite{lu2022deep}   & 0.737 / \textbf{0.851}  & 0.562 / 0.625 \\
\hline
$q_b$ (Ours) & 0.448 / 0.554 & 0.785 / 0.834 \\
$q_s$ (Ours)  & \textbf{0.837} / \textbf{0.891}  &  \textbf{0.805} / \textbf{0.844} \\
$q_t$ (Ours)  &  0.304 / 0.411 &  0.770 / 0.832 \\
$q_{st}$ (Ours)  &  0.787 / 0.831  & \textbf{0.807} / \textbf{0.843} \\

\end{tabular}

\caption{Performance comparison of our models against competing methods on BVI-SR and Waterloo-IVC-4K, with emphasis on spatial resolution changes. All numbers are presented in the SRCC / PLCC format. The top-$2$ results on each dataset are highlighted in \textbf{bold}.}
\label{performance_vaious_resolution}
\end{table}

\begin{table}\footnotesize
\centering
\renewcommand{\arraystretch}{1.25}
\begin{tabular}{lcc}
 Method  & BVI-HFR & LIVE-YT-HFR\\
\whline
NIQE~\cite{Mittal2013NIQE}  & 0.225 / 0.419 & 0.137 / 0.418 \\ 
BRISQUE~\cite{mittal2012no}  & 0.260 / 0.445 & 0.319 / 0.420 \\ 
VIDEVAL~\cite{tu2021ugc} &  0.345 / 0.474 & 0.475 / 0.567\\ 
TLVQM~\cite{korhonen2019two}    & 0.373 / 0.491 & 0.430 / 0.505\\ 
RAPIQUE~\cite{tu2021rapique}  & 0.304 / 0.463 & 0.457 / 0.567 \\ 
FAVER~\cite{zheng2022no}   & \textbf{0.556} / 0.639  &  0.635 / 0.692 \\
\hline

$q_b$ (Ours) &  0.035 / 0.206 &  0.588 / 0.696 \\
$q_s$ (Ours) & 0.194 / 0.460  & 0.584 / 0.690  \\
$q_t$ (Ours)  & \textbf{0.620} / \textbf{0.782} & \textbf{0.789} / \textbf{0.799} \\
$q_{st}$ (Ours)  &  0.541 / \textbf{0.685}  & \textbf{0.798} / \textbf{0.828}   \\

\end{tabular}

\caption{Performance comparison of our models against competing methods on BVI-HFR and LIVE-YT-HFR, with emphasis on frame rate changes.}
\label{performance_vaious_framerate}
\vspace{-12pt}
\end{table}

\begin{table}\footnotesize
\centering
\renewcommand{\arraystretch}{1.25}
\begin{tabular}{lcc}
 Method  & LIVE-Livestream & ETRI\\
\whline
NIQE~\cite{Mittal2013NIQE}  & 0.323 / 0.496 &  0.346 / 0.342 \\
BRISQUE~\cite{mittal2012no}  &  0.638 / 0.670 & 0.248 / 0.207 \\
TLVQM~\cite{korhonen2019two}    & \textbf{0.750} / \textbf{0.751} & 0.270 / 0.251 \\

\hline
$q_b$ (Ours) & 0.641 / 0.678 &  0.924 / 0.943 \\
$q_s$ (Ours)  & 0.669 / 0.720 &  0.926 / 0.944 \\
$q_t$ (Ours)  & 0.730 / 0.728  &  \textbf{0.929} / \textbf{0.947} \\
$q_{st}$ (Ours) & \textbf{0.802} / \textbf{0.811} &  \textbf{0.933} / \textbf{0.950} \\

\end{tabular}
\caption{Performance comparison of our models against competing methods on LIVE-Livestream and ETRI, with emphasis on video streaming-based artifacts.}
\label{performance_added_pgc}
\vspace{-12pt}
\end{table}

\begin{table*}\scriptsize
\centering
\renewcommand{\arraystretch}{1.25}
\resizebox{1\textwidth}{!}{
\begin{tabular}{l|c|c|c|c|c|c|c|c}
Method & CVD2014 & LIVE-Qualcomm & KoNViD-1K & LIVE-VQC & YouTube-UGC & LBVD & LIVE-YT-Gaming & \textit{Weighted Average} \\
\whline
VSFA~\cite{li2019quality}   & 0.850 / 0.869  & 0.708 / 0.774 &  0.794 / 0.799 &  0.718 / 0.771  & 0.787 / 0.789 & 0.834 / 0.830  &  0.784 / 0.819 & 0.789 / 0.805\\
Li22~\cite{li2022blindly}   & 0.863 / 0.883 &  \textbf{0.833} / \textbf{0.837} & 0.839 / 0.830 & 0.841 / 0.839  &   0.825 / 0.818 & \textbf{0.891} / \textbf{0.887}  &   0.852 / 0.868 & 0.849 / 0.847\\
SimpleVQA~\cite{sun2022deep}   & 0.834 / 0.864 & 0.722 / 0.774 &  0.792 / 0.798 & 0.740 / 0.775   & 0.819 / 0.817 & 0.872 / 0.878  & 0.814 / 0.836 & 0.810 / 0.823 \\
FastVQA~\cite{wu2022fast}   &  \textbf{0.883} / \textbf{0.901} &  0.807 / 0.814 &   \textbf{0.893} / 0.887 &    \textbf{0.853} / \textbf{0.873}  &  0.863 / 0.859 & 0.804 / 0.809   & \textbf{0.869} / 0.880 & 0.856 / 0.860\\
DOVER~\cite{wu2022disentangling}   & 0.858 / 0.881 & 0.736 / 0.789 & 0.892 / \textbf{0.900} &  \textbf{0.853} / 0.872  &    \textbf{0.875} / \textbf{0.874}  &  0.824 / 0.824  & \textbf{0.882} / \textbf{0.906} & 0.860 / 0.870 \\
\hline
$q_b$ (Ours) &  0.870 / 0.892 & 0.759 / 0.775 &  0.864 / 0.875 & 0.737 / 0.786     & 0.841 / 0.847 & 0.701 / 0.700     & 0.859 / 0.895 & 0.806 / 0.822 \\
$q_s$ (Ours)  & 0.873 / 0.892 & 0.804 / 0.806 & 0.868 / 0.878 &  0.714 / 0.776  & 0.857 / 0.859  & 0.678 / 0.683   & 0.857 / 0.898 & 0.805 / 0.822 \\
$q_t$ (Ours)  & 0.879 / 0.899 &  0.825 / 0.822  & 0.892 / 0.891 &  0.833 / 0.851 & 0.854 / 0.858 & 0.887 / 0.885  & 0.857 / 0.894 & \textbf{0.868} / \textbf{0.875} \\
$q_{st}$ (Ours)& \textbf{0.883} / \textbf{0.901} &  \textbf{0.832} / \textbf{0.842} & \textbf{0.901} / \textbf{0.905} & \textbf{0.860} / \textbf{0.880}  & \textbf{0.876} / \textbf{0.877}   & \textbf{0.898} / \textbf{0.892}  & 0.867 / \textbf{0.902} & \textbf{0.882} / \textbf{0.890}  \\

\end{tabular}
}
\caption{Performance comparison  of our models against five competing methods on seven small-scale UGC VQA datasets. The weighted average represents the average results across different datasets, weighted by the size of each dataset.}
\label{performance_ugc}

\end{table*}

\begin{table*}\footnotesize
\centering
\renewcommand{\arraystretch}{1.25}
\begin{tabular}{l|c|c|c|c|c|c|c}
\multirow{2}{*}{Method} & \multirow{2}{*}{\thead{Inference Time \\ (Sec)}} & \multicolumn{2}{c|}{In-dataset Testing} & \multicolumn{4}{c}{Cross-dataset Testing}\\

\cline{3-8}
   & & LSVQ-test & LSVQ-1080p & CVD2014 & KoNViD-1K & LIVE-VQC & YouTube-UGC \\
\whline
VSFA~\cite{li2019quality} & 11.109 & 0.801 / 0.796 & 0.675 / 0.704 & 0.756 / 0.760  & 0.810 / 0.811 & 0.753 / 0.795 & 0.718 / 0.721\\
Li22~\cite{li2022blindly} & 27.632 & 0.852 / 0.854 & 0.772 / 0.788 &  0.817 / 0.811 & 0.843 / 0.835  & 0.793 / 0.811  &\textbf{0.802} / 0.792 \\
SimpleVQA~\cite{sun2022deep} & 0.714 & 0.866 / 0.863 & 0.750 / 0.793 & 0.780 / 0.802 & 0.826 / 0.820 & 0.749 / 0.789 &  \textbf{0.802} / \textbf{0.806}   \\
FastVQA~\cite{wu2022fast} & 0.045 & 0.876 / 0.877 &  0.779 / 0.814 & 0.805 / 0.814 & 0.859 / 0.855 & \textbf{0.823} / \textbf{0.844} & 0.730 / 0.747  \\
DOVER~\cite{wu2022disentangling} & 0.047 & \textbf{0.888} / \textbf{0.889} & 0.795 / 0.830  & \textbf{0.829} / \textbf{0.832} & \textbf{0.884} / \textbf{0.883} & \textbf{0.832} / \textbf{0.855} & 0.777 / 0.792\\

\hline

$q_b$  (Ours) & 0.159 & 0.849 / 0.843 &  0.754 / 0.802 & 0.740 / 0.769 &  0.822 / 0.836 & 0.731 / 0.793 & 0.782 / 0.801\\
$q_s$ (Ours) & 0.159 & 0.838 / 0.842 & 0.764 / 0.808 & 0.775 / 0.796  & 0.845 / 0.856  & 0.773 / 0.823 & 0.723 / 0.743\\
$q_t$ (Ours) & 0.159  &  0.886 / 0.883 & \textbf{0.796} / \textbf{0.831} & 0.816 / 0.837 & 0.851 / 0.853 & 0.803 / 0.837 & 0.774 / 0.791\\
$q_{st}$ (Ours) & 0.159 &  \textbf{0.895} / \textbf{0.895} & \textbf{0.809} / \textbf{0.844}  & \textbf{0.823} / \textbf{0.839} & \textbf{0.878} / \textbf{0.884} & 0.806 / \textbf{0.844} &  0.788 / \textbf{0.804}\\

\end{tabular}

\caption{In-dataset and cross-dataset testing of our model against five competing models, all retrained on the official training split of the large-scale LSVQ dataset~\cite{ying2021patch} and tested on other VQA datasets without fine-tuning. Inference time is computed using an NVIDIA A100 GPU, on an $8$-second, $1080$p, and $30$ fps video.
}
\label{cross_valid}
\vspace{-9pt}
\end{table*}

\noindent\textbf{Spatial Rectifier Matters in Responding to Spatial Resolution Changes}. This is evident in Table~\ref{performance_vaious_resolution}, especially on the BVI-SR dataset~\cite{mackin2018studyreso}. By incorporating the spatial rectifier, our model 
shows substantial improvements in performance compared to all other methods. Furthermore, it is worth noting that the inclusion of the temporal rectifier does not contribute significantly and can even harm the accuracy of predictions
concerning distortions related to spatial resolution changes. This underscores the value of a modular approach in the construction of BVQA models. 

Nevertheless, we notice that the base quality predictor, despite working with low-spatial-resolution input, performs far better on Waterloo-IVC-4K~\cite{li2019avc} than BVI-SR. This performance discrepancy is largely attributed to the dominant compression artifacts present in  Waterloo-IVC-4K, which are relatively unaffected by changes in spatial resolution. Conversely, BVI-SR is specifically designed to address distortions resulting from spatial subsampling.

Additionally, even though NIQE, BRISQUE, and VSFA extract global frame features at the actual spatial resolution, they show subpar performance on Waterloo-IVC-4K.  Similarly, relying on cropped patches (\ie, local frame features), BTURA also shows limited effectiveness on this dataset. These outcomes indicate that simple global or complex local feature extraction is inadequate when confronted with videos compressed at varying spatial resolutions.

\noindent\textbf{Temporal Rectifier Matters in Responding to Frame Rate Changes}. This is evident in Table~\ref{performance_vaious_framerate}, especially on BVI-HFR~\cite{mackin2018studyfr}. The inclusion of the temporal rectifier significantly enhances the performance of our model, surpassing all alternative methods by a clear margin. We find that the spatial rectifier tends to be effective primarily when combined with the temporal rectifier, as in the complete model, $q_{st}$. Similarly, the base quality predictor delivers noticeably better performance on LIVE-YT-HFR~\cite{madhusudana2021subjective} compared to BVI-HFR. This difference can be ascribed to the fact that LIVE-YT-HFR contains a certain portion of spatial compression distortions, which the base quality predictor is adept at quantifying.

Out of the six competing models in Table~\ref{performance_vaious_framerate}, only FAVER~\cite{zheng2022no} is specifically engineered to assess quality fluctuations due to frame rate changes, and it performs the best among them.  Nevertheless, its subpar performance relative to our model suggests that the handcrafted features in FAVER have limited capability in addressing distortions related to frame rate variations.

\noindent\textbf{Full Model Performs the Best in Complex Video Streaming Applications}. LIVE-Livestream~\cite{shang2021assessment} is subject to both streaming and compression distortions,  whereas ETRI~\cite{lee2021subjective} is affected by spatiotemporal subsampling due to different levels of compression, primarily for streaming applications. The outcomes in Table~\ref{performance_added_pgc} demonstrate that incorporating both spatial and temporal rectifiers is crucial for optimal performance on both datasets.

\subsection{Results on UGC Datasets}
Our method is pretrained on the official training split of the large-scale LSVQ dataset~\cite{ying2021patch}, followed by fine-tuning on seven small-scale UGC datasets~\cite{nuutinen2016cvd2014,ghadiyaram2017capture,hosu2017konstanz,sinno2018large,wang2019youtube,chen2019qoe,yu2022subjective}. All competing models are retrained following the procedures outlined in their respective papers. Results are shown in Table~\ref{performance_ugc}. 
Furthermore, a cross-dataset testing is performed, with results shown in Table~\ref{cross_valid}.

\begin{table*}[t]
\vspace{-.2em}
\centering
\subfloat[
\textbf{Backbone initialization}. ViT-B outperforms ResNet-50, and 
 the initialization using contrastive learning in CLIP is favored over the supervised learning on ImageNet-1K.
\label{tab:backbone}
]{
\centering
\begin{minipage}{0.29\linewidth}{\begin{center}
\tablestyle{1.3pt}{1.1}
\begin{tabular}{y{54}x{42}x{44}}
     & LSVQ-test & LSVQ-1080p  \\
    \whline
    RN50-ImageNet & 0.858 / 0.857 & 0.741 / 0.782 \\
    RN50-CLIP & 0.870 / 0.870 & 0.780 / 0.812\\
    ViT-ImageNet & 0.866 / 0.867 & 0.758 / 0.801\\
    ViT-CLIP& \baseline{\textbf{0.895} / \textbf{0.895}} & \baseline{\textbf{0.809} / \textbf{0.844}} \\
\end{tabular}
\end{center}}\end{minipage}
}
\hspace{2em}
\subfloat[
\textbf{Spatial resizing}. NN stands for the nearest neighbor interpolation. Bicubic interpolation marginally outperforms bilinear and nearest neighbor interpolation.
\label{tab:resize} 
]{
\centering
\begin{minipage}{0.29\linewidth}{\begin{center}
\tablestyle{3.0pt}{1.1}
\begin{tabular}{y{25}x{45}x{45}}
  & LSVQ-test & LSVQ-1080p \\
\whline
NN &  0.893 / 0.893  & 0.809 / 0.844 \\ 
Bilinear  &  0.891 / 0.891  & 0.809 / 0.843 \\
Bicubic &  \baseline{\textbf{0.895} / \textbf{0.895}}  &  \baseline{\textbf{0.809} / \textbf{0.844}} \\
 &    &  \\
\end{tabular}
\end{center}}\end{minipage}
}
\hspace{2em}
\vspace{0.3cm}
\subfloat[
\textbf{Regressor}. GRU indicates the gated recurrent unit~\cite{cho2014learning}. Even in the absence of complex temporal modeling, MLP exhibits a slight advantage over GRU and Transformer models.
\label{tab:regressor}
]{
\centering
\begin{minipage}{0.29\linewidth}{\begin{center}
\tablestyle{3pt}{1.1}
\begin{tabular}{y{36}x{42}x{44}}
  & LSVQ-test & LSVQ-1080p \\
\whline
MLP &  \baseline{\textbf{0.895} / \textbf{0.895}}  & \baseline{\textbf{0.809} / \textbf{0.844}} \\
GRU &    0.894 / 0.895 & 0.808 / 0.844 \\
Transformer &  0.892 / 0.891 & 0.802 / 0.841 \\
 &    &  \\
\end{tabular}
\end{center}}\end{minipage}
}
\hspace{2em}
\subfloat[
\textbf{Loss function}. PLCC proves to be significantly better than the $\ell_1$-norm and $\ell_2$-norm induced metrics as loss functions.
\label{tab:loss}
]{
\centering
\begin{minipage}{0.29\linewidth}{\begin{center}
\tablestyle{3pt}{1.1}
\begin{tabular}{y{20}x{45}x{45}}
  & LSVQ-test & LSVQ-1080p \\
\whline
$\ell_1$ &   0.882 / 0.883 & 0.779 / 0.823 \\
$\ell_2$ &    0.887 / 0.886 &  0.797 / 0.828 \\
PLCC &   \baseline{\textbf{0.895} / \textbf{0.895}}  &  \baseline{\textbf{0.809} / \textbf{0.844}} \\
\end{tabular}
\end{center}}\end{minipage}
}
\hspace{2em}
\subfloat[
\textbf{Spatial rectifier}. As a shallower architecture with fewer parameters, ResNet-18 is sufficient to achieve satisfactory results.
\label{tab:spatial}
]{
\centering
\begin{minipage}{0.29\linewidth}{\begin{center}
\tablestyle{3pt}{1.1}
\begin{tabular}{y{20}x{43}x{60}}
  & BVI-SR & Waterloo-IVC-4K  \\
\whline
RN18 &  \baseline{0.859 / 0.904} & \baseline{\textbf{0.850} / \textbf{0.870}}  \\
RN34 &   \textbf{0.866} / \textbf{0.907} & 0.810 / 0.855 \\
RN50 &   0.861 / 0.911 & 0.809 / 0.846  \\
\end{tabular}
\end{center}}\end{minipage}
}
\hspace{2em}
\subfloat[
\textbf{Temporal rectifier}. S and T stand for the slow and fast pathways in SlowFast~\cite{feichtenhofer2019slowfast}. The fast pathway solely yields excellent performance with reduced complexity.
\label{tab:temporal}
]{
\centering
\begin{minipage}{0.29\linewidth}{\begin{center}
\tablestyle{2.0pt}{1.1}
\begin{tabular}{y{15}x{43}x{50}}
 & BVI-HFR & LIVE-YT-HFR  \\
\whline
S &  0.423 / 0.478  &  0.515 / 0.629   \\
T &  \baseline{\textbf{0.629} / \textbf{0.733}} &  \baseline{\textbf{0.791} / \textbf{0.801}}  \\
S+T &  0.529 / 0.659 &  0.765 / 0.789  \\
\end{tabular}
\end{center}}\end{minipage}
}
\hspace{2em}
\vspace{-.2em}
\caption{Ablation experiments. Default settings are marked in \colorbox{baselinecolor}{gray}.}
\label{table:ablations} 
\vspace{-9pt}
\end{table*}

\noindent\textbf{Base Quality Predictor Works Well on CVD2014~\cite{nuutinen2016cvd2014}, KoNViD-1K~\cite{hosu2017konstanz}, and LIVE-YT-Gaming~\cite{yu2022subjective}}. The base quality predictor achieves commendable performance on CVD2014. Akin to a blind image quality model, it outperforms most alternatives, except for FastVQA~\cite{wu2022fast}. Incorporating the spatial and temporal rectifiers yields marginal performance gains. These results provide a strong indication that CVD2014 is primarily influenced by spatial distortions that are insensitive to spatial resizing with limited content diversity (\ie, composition of merely five natural scenes). Similar observations can be made on  KoNViD-1K and LIVE-YT-Gaming, which are dominated by spatial distortions with some spatial complexity.

\noindent\textbf{Spatial Rectifier Matters on YouTube-UGC~\cite{wang2019youtube} and LIVE-Qualcomm~\cite{ghadiyaram2017capture}}. 
The spatial rectifier offers favorable performance improvements on  YouTube-UGC and LIVE-Qualcomm.
This can be explained by the presence of ultra-high-spatial-resolution videos, such as $4$K in YouTube-UGC and  $1080$p  in LIVE-Qualcomm. Similar conclusions can be drawn by comparing $q_b$ with $q_s$ on LSVQ-1080p. 

\noindent\textbf{Temporal Rectifier Matters on LIVE-VQC~\cite{sinno2018large}, LIVE-Qualcomm~\cite{ghadiyaram2017capture}, and LBVD~\cite{chen2019qoe}}. The temporal rectifier proves beneficial across multiple datasets, particularly on LIVE-VQC, LIVE-Qualcomm, and LBVD. This suggests the prominence of temporal distortions in these datasets. Interestingly, the spatially rectified model $q_s$ is inferior to the base model $q_b$ on LIVE-VQC and LBVD, providing additional evidence of the dominance of temporal distortions.

\noindent\textbf{Full Model Performs the Best with Strong Cross-Dataset Generalization}. The combination of both rectifiers consistently improves the quality prediction performance across all eight UGC datasets, yielding results that rival those of the leading methods. More importantly, the fully rectified model exhibits strong cross-dataset generalization, affirming the effectiveness of our modular BVQA method in addressing spatial and temporal aspects of video quality distinctly and thoroughly.

Our computational analysis of the eight UGC datasets resonates with a recent computational investigation of VQA datasets via design of minimalistic VQA models~\cite{sun2024analysis}. Specifically, we also expose the \textit{easy dataset problem} in the construction of current VQA datasets, which allows over-simplistic models, such as those close to blind image quality models, to achieve state-of-the-art performance. The root cause of this problem is the superficial treatment of sample selection and excessive reliance on absolute category rating as the de facto subjective testing methodology to gather single-valued quality indicators.

\subsection{Ablation Studies}
We expand our computational investigation by ablating 1) backbones of the based quality predictor with different initializations in Table~\ref{tab:backbone}, 2) spatial resizing methods in Table~\ref{tab:resize}, 3) quality regressors in Table~\ref{tab:regressor}, 4) loss functions in Table~\ref{tab:loss}, 5) spatial rectifiers in Table~\ref{tab:spatial}, and 6) temporal rectifiers in Table~\ref{tab:temporal}. For the former four studies, we ablate the full model $q_{st}$ on LSVQ~\cite{ying2021patch}, for the spatial rectifier, we ablate $q_s$ on BVI-SR~\cite{mackin2018studyreso} and Waterloo-IVC-4K~\cite{li2019avc}, and for the temporal rectifier, we ablate $q_t$  on BVI-HFR~\cite{mackin2018studyfr} and LIVE-YT-HFR~\cite{madhusudana2021subjective}.

The ablation studies show that our base quality predictor benefits from a stronger ViT-B~\cite{dosovitskiy2020image} backbone, pretrained through contrastive learning. The impact of employing different spatial resizing techniques for preparing inputs to the base predictor and the two rectifiers appears to be negligible, so do the quality regressors. 
Even more sophisticated models like the gated recurrent unit (GRU)~\cite{cho2014learning} and the Transformer model~\cite{vaswani2017attention} do not show expected performance improvements over the MLP, when considering the increased computational complexity. 
In contrast,  PLCC as the loss function yields noticeable performance gains compared to the $\ell_1$-norm and $\ell_2$-norm induced metrics. Moreover, the 
simple ResNet-18 as the spatial rectifier and the Fast pathway as the temporal rectifier provide a good trade-off between performance and computational complexity.

\section{Conclusion and Discussion}
We have described a modular BVQA model, accompanied by a training strategy that improves its modularity. Our method addresses the BVQA challenges posed by intricacies of natural scenes and spatiotemporal distortions, and variations in spatial resolutions and frame rates. Extensive experiments on fourteen VQA datasets verify the promise of the proposed method. Additionally, the modular design of our model gives us an opportunity to analyze current UGC datasets in terms of their spatial and temporal complexity.

The current work serves as an initial step towards modular BVQA, yet numerous challenges remain unaddressed. For example, our method assumes a fixed routing function~\cite{pfeiffer2023modular}, where activation of the rectifiers is determined by expert insight into the VQA datasets. A routing function that adapts to specific datasets or individual samples is compelling for future research. Meanwhile, the handcrafted aggregation function in Eq.~\eqref{eq:2_sr_tr} might seem rudimentary. It is interesting to learn a neural network to aggregate active modules, potentially at the feature level. Finally, adding a new rectifier currently necessitates the retraining or fine-tuning of all parameters. Exploring parameter-efficient training methodologies, such as continual learning~\cite{zhang2022continual} and transfer learning, could offer substantial benefits.

\section{Acknowledgement}

This work was supported in part by the National Natural Science Foundation of China under Grants 62071407 and 62102339,  and the
Hong Kong RGC Early Career Scheme (2121382).

{
    \small
    \bibliographystyle{ieeenat_fullname}
    \bibliography{main}
}

\end{document}